# The expressway network design problem for multiple urban subregions based on the macroscopic fundamental diagram

Yunran Di[a], Weihua Zhang[b], Haotian Shi[c,*], Heng Ding[b,*], Jinbiao Huo[d], Bin Ran[c]

[a]College of Civil Engineering, Hefei University of Technology, Hefei 230009, China.
[b]School of Automotive and Traffic Engineering, Hefei University of Technology, Hefei 230009, China.
[c]Department Civil and Environmental Engineering, University of Wisconsin-Madison, WI 53706, USA.
[d]School of Transportation, Southeast University, Nanjing, China.

Yunran Di, Ph.D. Candidate, Email: diyunran@mail.hfut.edu.cn
Weihua Zhang, Professor, Email: weihuazhang@hfut.edu.cn
Haotian Shi, Research Associate, Email: hshi84@wisc.edu
Heng Ding, Professor, Email: dingheng@hfut.edu.cn
Jinbiao Huo, Ph.D. Candidate, Email: jinbiaoh@outlook.com
Bin Ran, Professor, Email: bran@wisc.edu

Corresponding authors: Heng Ding (dingheng@hfut.edu.cn) and Haotian Shi (hshi84@wisc.edu).

**Acknowledgements:** This research was financially supported by the National Natural Science Foundation of China (Grant Nos. 52072108 and 52372326), the Municipal Natural Science Foundation of Hefei (Grant No. 2022020) and China Scholarship Council Awards.

**Declaration of interests**
The authors declare that they have no known competing financial interests or personal relationships that could have appeared to influence the work reported in this paper.

# The expressway network design problem for multiple urban subregions based on the macroscopic fundamental diagram

**Abstract:** As urbanization advances, cities are expanding, leading to a more decentralized urban structure and longer average commuting durations. The construction of an urban expressway system emerges as a critical strategy to tackle this challenge. However, the traditional link-level network design method faces modeling and solution challenges when dealing with the large-scale expressway network design problem (ENDP). To address the challenges, this paper proposes an expressway network design method for multiple urban subregions based on the macroscopic fundamental diagram (MFD). Initially, a mixed road network traffic model that describes traffic dynamics of multiple subregions and candidate expressways is developed by integrating the MFD and the cell transmission model (CTM). Then, treating urban subregions and candidate expressways as route nodes in the mixed road network, a route choice model is established based on stochastic user equilibrium. Finally, a decision model for ENDP is proposed to minimize vehicle travel time under the construction budget constraint. The impact of financial investment and traffic demand on expressway network design schemes in the case study is explored separately. The simulation results indicate that during the initial stages of expressway planning, the construction of new expressways can significantly alleviate traffic congestion. However, as the expressway network expands further, the effectiveness of improving traffic conditions through new expressway construction gradually diminishes if traffic demand does not continue to increase. Additionally, variations in traffic demand between subregions result in different construction schemes, emphasizing the importance of adjusting budget allocations based on specific traffic demands.

## 1. Introduction

According to the World Cities Report 2022 by the United Nations, urban areas already accommodate 55% of the global population, a figure projected to rise to 68% by 2050 (UN-Habitat, 2022). This rapid urbanization results in a surge in urban travel demand, particularly for long-distance journeys. Consequently, there is an urgent necessity for new transportation infrastructure and enhancements to existing structures to meet the needs of the expanding population and increasingly intricate travel patterns within urban regions (Iacono and Levinson, 2009). As high-level road types within the transportation system, expressways offer long-distance and high-speed transit services and influence urban sprawl and trip distribution (Kim, 2007; Zhu et al., 2023). The design of expressways is critical for travel decisions. For instance, in Beijing, major urban expressways, comprising only 8% of the total road network length, handle nearly 50% of the traffic flow; similarly, in Shanghai, urban expressways, covering just 5% of the road network, carry over 35% of the city's traffic volume (Zhao et al., 2009; Lu et al., 2011). Compared to standard roads, expressways feature higher design speeds with no intersections, serving as fast channels for inter-regional traffic transmission. Consequently, urban expressways have become integral to modern urban transportation networks (Clevenger et al., 2013; Hashemi et al., 2021).

Transitioning from the context of urban expressway expansion to strategic transportation planning, it becomes evident that the increasing demand for efficient transportation necessitates optimizing network design. In transportation network modeling and optimization, the modification of a transportation network configuration by adding new links or improving existing ones is referred to as the road network design problem (RNDP) (Farvaresh and Sepehri, 2011). RNDPs are typically framed as bi-level programming, wherein all variants strive to minimize (or maximize) specific network performance metrics (e.g., total travel time or generalized cost) at the upper layer while simultaneously accounting for travelers' route choice behavior in a user equilibrium (UE) or stochastic user equilibrium (SUE) manner at the lower layer (Yang and Bell, 2001; Chiou, 2024). Various effective solution techniques have been developed for solving RNDP optimization objectives, such as branch and bound method, simulated annealing, and genetic algorithm (Farahani et al., 2013).

Depending on the nature of the decision variables, RNDPs are categorized into discrete network design problems (DNDPs), continuous network design problems (CNDPs), and mixed network design problems (MNDPs). DNDPs focus on determining the addition of links to an existing road network to minimize total system travel time (Farvaresh and Sepehri, 2011; Wang et al., 2013; Haas and Bekhor, 2017; Wang et al., 2015). For instance, Farvaresh and Sepehri (2011) proposed a single-level mixed integer linear programming formulation for bi-level DNDP to offer optimal network design solutions for a small square road network. Wang et al. (2013) introduced a bi-level programming model grounded on DNDP to ascertain the optimal number of lanes for each candidate link in a road network by minimizing the upper layer objective of total travel time and addressing the lower layer issue of Wardrop user equilibrium. CNDPs are concerned with continuous design decisions, such as expanding the capacity of streets and determining tolls for some specific streets. Wang et al. (2014) proposed a mathematical programming model with equilibrium constraints to solve the optimal link capacity expansion problem in a road network. Fan and Gurmu (2014) ascertained the optimal solution to the integrated congestion pricing and capacity expansion problem using a bi-level genetic algorithm. MNDPs are NDPs that make both continuous and discrete decisions. For example, Szeto and Jiang et al. (2015) proposed a model that combines road link direction redesign, road capacity expansion, and intersection signal settings. Further literature on RNDP can be found in review papers authored by Farahani et al. (2013), Xu et al. (2016), and Jia et al. (2019).

As research progresses from isolated points to the basic units of a community and eventually to the entire network, it has been consistently observed that the traffic dynamics of urban networks often exhibit a macroscopic fundamental diagram (MFD). The MFD delineates the relationship between vehicle accumulation and macroscopic characteristics of a homogeneous network, such as average speed and outflow rate, and has become a beneficial tool for assessing macroscopic traffic flow dynamics. Many congestion mitigation strategies and policies have been developed based on MFD dynamic models, including perimeter control (Haddad and Mirkin, 2020; Ding et al., 2022), congestion pricing (Zheng et al., 2016; Gu et al., 2018), route guidance (Ding et al., 2017; Sirmatel and Geroliminis, 2017; Di et al., 2024), and regional traffic assignment (Batista and Leclercq, 2019; Yildirimoglu and Geroliminis, 2014). In recent years, MFD and its extended multimodal MFD have gained increasing significance in evaluating urban road network design. For instance, considering that modifying road infrastructure or signal control can impact MFD, Hu et al. (2020) established a DNDP using the MFD of the target road network as a performance indicator. Zheng et al. (2017) proposed a method to optimize the spatial allocation of bus lanes in an urban network by describing multimodal traffic dynamics and travel costs using multimodal MFD. Dakic et al. (2021) investigated the optimal design problem of bus networks based on multimodal MFD.

Despite many efforts on road network optimization, not enough attention has been paid to the design and planning of expressways. In the few studies available, Santos et al. (2008) incorporated the concept of fairness into road design schemes and proposed a network design approach for expressways. Additionally, Di et al. (2018) tested a proposed traffic network design method using an expressway network. However, these studies treat expressways as separate entities and do not meet realistic requirements. Waller and Ziliaskopoulos (2001) modeled urban arterials and expressways using the cell transmission model (CTM) to investigate road design issues. Nevertheless, the study examines the traffic operations of individual road links. Due to its computational complexity, this link-based modeling method is more suitable for describing lane dynamics within small-scale road networks. Recently, some studies have focused on lane design issues within expressways. For example, Movaghar et al. (2020) and Chakraborty et al. (2021) proposed methods for determining dedicated lanes for autonomous vehicles in expressway network design, but these methods cannot be used to determine whether expressways should be constructed. In summary, existing studies on expressway design exhibit the following gaps: firstly, a single modeling approach struggles to accurately describe the traffic characteristics of mixed road networks comprising both expressways and arterial roads. Secondly, the significant impact of traffic demand on the urban road network on expressway design has not been adequately considered.

Based on the above discussions, it is clear that expressways are not isolated entities within urban networks; thus, the presence of arterial networks must be integrated into expressway planning. This integration presents challenges in addressing the expressway network design problem (ENDP). Compared to standard roads, expressways serve as the main arteries of urban networks, typically featuring higher transportation capacities and longer planned lengths (Li et al., 2019). Consequently, when designing expressways, it is essential to construct traffic models for large-scale urban networks comprising numerous nodes and links. Existing RNDP methods primarily focus on specific networks composed of intersections and links, addressing design issues of limited-scale networks by analyzing traffic transmission patterns on links. Due to the large-scale nature of the urban networks involved in expressway planning, traditional RNDP methods face difficulties in modeling and solving the optimization objectives for ENDP.

Inspired by the application of the MFD in RNDPs, we innovatively introduce the concept of MFD into ENDP to address this challenge and take the following measures to address the above-identified gaps. First, the MFD model is considered to describe the macroscopic traffic state of the arterial network in ENDP to avoid the microscopic description of numerous arterial lanes and reduce computational costs. Meanwhile, CTM is used to model expressways to capture the characteristics of expressways as links between subregions. This combination of CTM and MFD establishes a macroscopic traffic transfer mechanism between expressways and arterial networks, facilitating the construction of a comprehensive traffic model for the mixed road network. Furthermore, since the role of urban expressways is to transport traffic from arterial networks, it is necessary for expressway construction to accommodate the traffic demand distribution between different urban areas. To this end, this study designs a multi-subregion network scenario to establish an ENDP decision model between subregions based on their spatial relationships and traffic demand.

To summarize, this paper proposes an expressway network design problem for multiple urban subregions based on the MFD. The study encompasses two main parts: 1) constructing the traffic dynamics of the mixed road network and 2) establishing a decision model for ENDP, where traffic dynamics serve as the foundation for network design decisions. For traffic dynamics modeling, the MFD and CTM are used to describe the traffic states of the arterial network and candidate expressways, respectively. Subsequently, a traffic model of the mixed road network is integrated by developing the traffic flow exchange mechanism between the expressway and the connecting subregions. In this way, a route selection model based on the more realistic SUE concept is devised, thus constructing the traffic dynamics of the mixed road network. For expressway network design decisions, this study proposes a decision model in the context of urban networks with multiple subregion structures. This decision model takes whether to build candidate expressways as decision variables and aims to minimize the total travel time (TTS) of all travelers while adhering to investment budget constraints. With the traffic demand between subregions as input, the optimization problem is solved to determine which subregion's expressways should be constructed.

This study makes two main contributions to the mixed road network modeling and network design problems. (1) Regarding modeling, we integrate the traffic dynamics of arterial networks and expressway systems using both region-based and link-based models. This approach simplifies the modeling process for arterial networks while preserving the attributes of expressways as link-level corridors between regions. In addition, it also facilitates the construction of route choice models for mixed road networks at the macroscopic level. (2) Regarding the network design problem, this study places the planning of expressways in a multi-subregion scenario, which enables practitioners to consider the impact of traffic demand distribution between different urban areas in the expressway network design process.

The remaining sections are summarized below. Section 2 describes the research problem. Section 3 develops the traffic dynamics for the mixed road network. Section 4 proposes the expressway network design problem, and Section 5 presents the results of the case study. In the last section, conclusions are drawn, and future work is summarized.

## 2. Problem description

### 2.1 Scenario description and assumptions

Consider an urban network partitioned into several connected subregions. Based on the existing arterial road network, there are plans to construct several new expressways between subregions as alternative travel routes. Each candidate expressway has two travel directions, and the two opposite travel directions do not directly exchange traffic due to physical separation. Therefore, the expressway is treated separately according to the travel direction, with the expressway traveling from subregion $i$ to subregion $j$ denoted as $E_{ij}, i \neq j$.

In planning expressways, determining site selection and ramp layout is complex, requiring consideration of various factors such as the urban network structure, environmental impacts, and societal factors. This paper focuses on expressway planning from a macroscopic perspective, emphasizing traffic demand. Given the numerous parameters involved in expressway planning, including the number and distribution of ramps, the following assumptions are made to streamline the structure and route selection method of the mixed road network:

(i) Each candidate expressway is equipped with ramps only in the origin and destination subregions. In other words, $E_{ij}$ (if it exists) has only one on-ramp in subregion $i$ and one off-ramp in subregion $j$. Denote the on-ramp of expressway $E_{ij}$ in subregion $i$ as $r_{iE_{ij}}$ and the off-ramp of expressway $E_{ij}$ in subregion $j$ as $r_{E_{ij}j}$.

(ii) The endpoints of all expressways within the same subregion are situated within a particular localized area. Given a subregion $h \neq i, j$, if there is an expressway $E_{hi}$ that intersects with expressway $E_{ij}$ within subregion $i$, the connecting ramp $r_{E_{hi}E_{ij}}$ will be installed to allow traffic flow from expressway $E_{hi}$ to expressway $E_{ij}$.

(iii) Vehicles have a maximum of one transfer within the same subregion (expressway) during each trip, i.e., vehicles are not allowed to leave a subregion (expressway) and then return to it again.

Assumptions (i) and (ii) entail settings for the road network structure, while assumption (iii) specifies rules for travel routes. Here, further elaboration on these assumptions is provided. Regarding (i), although scenarios with multiple ramps for a single expressway are not considered, this assumption still reflects the influence of constructing expressways on inter-subregion travel patterns, as for (ii), it is a realistic setting. We elucidate the rationality behind this by referring to the scenario depicted in Fig. 1. Fig. 1 illustrates a mixed road network comprising five subregions and four expressways, where all four expressways connect to subregion 5. In Fig. 1(a), multiple expressways pass through subregion 5, but the endpoints within subregion 5 are far apart, indicating a lack of direct connections. Consequently, traffic transitions between different expressways within the same subregion must occur via the arterial network, increasing route complexity and reducing expressway network connectivity. Based on (ii), Fig. 1(b) presents an improved road network structure, aligning more closely with expressway design requirements. Additionally, the settings in (iii) adhere to common travel rules.

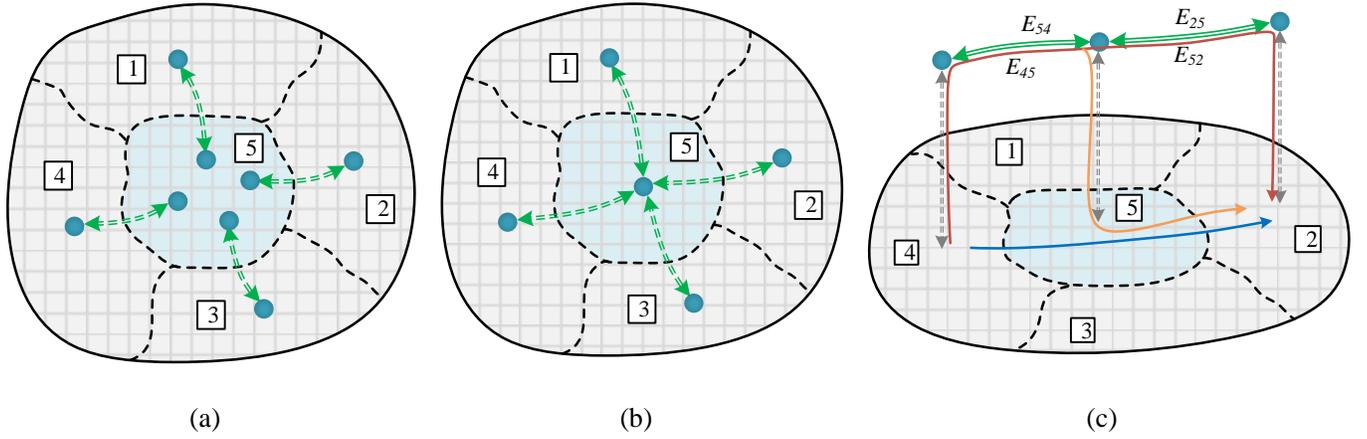

(a) (b) (c)

Fig. 1 Layout pattern of expressways between subregions.

The introduction and layout of expressways have brought about significant changes to the original road network structure, resulting in a notable shift in travel patterns. In the absence of expressways, traffic between adjacent subregions can only transfer through the passageway located at the boundary. Let $r_{ij}$ denote the boundary passageway that travels in a single direction from subregion $i$ to neighboring subregion $j$. With the presence of expressways, route selection becomes more diverse. Taking the travel demand from subregion 4 to subregion 2 as an example, Fig. 1(c) illustrates three different travel routes: route 4→5→2, route 4→$E_{45}$→$E_{52}$→2, and route 4→$E_{45}$→5→2.

**2.2 Research framework**

Based on the problem descriptions in Section 2.1, the mixed road network can be regarded as coupling an arterial road network and an expressway network. Within the mixed road network, every subregion and expressway is treated as a node along the route, forming a network of numerous nodes and links. The structure of the mixed road network is represented by $G = (\mathbb{R}, \mathbb{N}, \mathbb{C})$, where $\mathbb{R}$ denotes the set of subregions $\mathbb{R} = \{1,...,i_m\}$, and $i_m$ is the total number of subregions; $\mathbb{N}$ denotes the set of potential candidate expressways, $\mathbb{N} = \{E_{ij}\}$, $i, j \leq i_m, i \neq j$; and $\mathbb{C}$ represents the set of four types of links between nodes, $\mathbb{C} = \{r_{ij}, r_{iE_{ij}}, r_{E_{hi}i}, r_{E_{hi}E_{ij}}\}$.

The ENDP addressed in this paper can be viewed as planning an expressway network within the existing topology of the arterial road network. The core of this study lies in determining which subregions should construct new expressways under budget constraints to enhance the traffic efficiency of the road network. Fig. 2 illustrates the research framework for ENDP in a multi-subregion scenario. Within this framework, two main issues need to be addressed. Firstly, we establish the mixed road network traffic dynamics to resolve the lower-layer traffic assignment problem. The approach involves establishing subregion and expressway network traffic models based on the MFD and CTM, forming the mixed road network traffic model. A route choice model for the mixed road network is developed to simulate traffic dynamics. Secondly, constructing an optimization model for expressways to address upper-layer network design problems. Considering the multi-subregion network structure and budget constraints, we propose an optimization model to minimize the TTS of all travelers. This model determines the optimal design scheme for expressways between subregions. The details of the methodology are introduced in the subsequent sections.

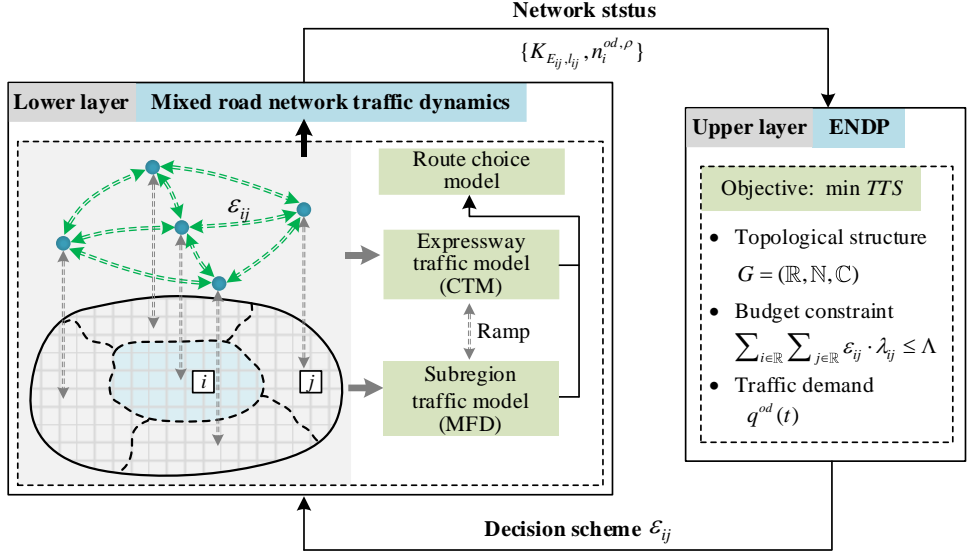

Fig. 2 Research framework.

## 3. Traffic dynamics modeling for mixed road networks

This section conducts traffic modeling separately for both subregions and candidate expressways, and based on this, develops a route choice model for the mixed road network. The established traffic dynamic model will be utilized to observe the traffic status of mixed road networks. For ease of distinction, define $\alpha_i$ as the set of immediately neighboring subregions of subregion $i$ ($\alpha_i$ is a non-empty set), and $\beta_i$ as the set of expressways directly connected to subregion $i$.

### 3.1 Subregion traffic model

The MFD provides a single-peak, low-scattering relationship between the accumulation of vehicles and the spatially averaged speed or production for a homogeneously congested arterial network. Consider a homogeneous subregion $i \in \mathbb{R}$ with travel production MFD $P_i(n_i(t)) = v_i(t) \cdot n_i(t)$, where $n_i(t)$ and $v_i(t)$ are the accumulation of vehicles and the average travel speed in subregion $i$ at time $t$, respectively.

Each route in the MFD is defined as a sequence of nodes from an origin subregion to a destination subregion. A route $\rho$ is defined as $\rho = \{\rho_1, \rho_2, ..., \rho_{k_\rho}\}$, where $k_\rho$ is the number of nodes. Since the traffic demand originates and terminates within the arterial road network, $\rho_1$ and $\rho_{k_\rho}$ must be subregions, while the other nodes can be either subregions or expressways. Let $n_i^{od,\rho}(t)$ denote the number of vehicles in subregion $i$ at time $t$ with the origin subregion $o$, destination subregion $d$, and route $\rho$. Thus, $n_i(t) = \sum_o \sum_d \sum_\rho n_i^{od,\rho}(t)$. Further, let $n_{i \to \rho(+i)}^{od,\rho}(t)$ denote the number of vehicles traveling to node $\rho(+i)$ in subregion $i$ at time $t$, where $\rho(+i)$ is the next node of subregion $i$ in route $\rho$. Let $\rho(-i)$ denote the previous node of subregion $i$ in route $\rho$. The trip completion rate for vehicles to the next node $\rho(+i)$ in subregion $i$ at time $t$ with origin subregion $o$ and destination subregion $d$ is:

$$m_{i \to \rho(+i)}^{od,\rho}(t) = \frac{n_{i \to \rho(+i)}^{od,\rho}(t)}{n_i(t)} \cdot \frac{P_i(n_i(t))}{\bar{L}_i} \tag{1}$$

where $\bar{L}_i$ denotes the average trip length of the vehicles in subregion $i$.

For a subregion, the traffic demand consists of two parts: the transfer flow from connected subregions or expressways and the newly generated exogenous traffic demand within this subregion. Let $q^{od}(t)$ denote the newly generated exogenous traffic demand from subregion $o$ to subregion $d$ at time t, and $q^{od,\rho}(t)$ be the demand assigned to route $\rho$,

$q^{od}(t) = \sum_{\rho} q^{od,\rho}(t)$. The traffic assignment results are estimated by the logit route choice formula (Moshahedi and Kattan, 2023), which will be presented in Section 3.3. The dynamics of a subregion are due to the difference between its inflows (both initiated and transferred trips) and outflows (both terminated and transferred trips). According to the principle of accumulation conservation, the dynamic equation of traffic in subregion $i$ for route $\rho$ is:

$$\frac{dn_i^{od,\rho}(t)}{d(t)} = \begin{cases} q^{od,\rho}(t) - m_{i \to i}^{od,\rho}(t) & \text{(i) if } i = o \,\&\, i = d \\ q^{od,\rho}(t) - \hat{m}_{i \to \rho(+i)}^{od,\rho}(t) & \text{(ii) if } i = o \,\&\, i \neq d \\ y_{\rho(-i) \to i}^{od,\rho}(t) - m_{i \to \rho(+i)}^{od,\rho}(t) & \text{(iii) if } i \neq o \,\&\, i = d \\ y_{\rho(-i) \to i}^{od,\rho}(t) - \hat{m}_{i \to \rho(+i)}^{od,\rho}(t) & \text{(iv) otherwise} \end{cases} \quad (2)$$

where $m_{i \to i}^{od,\rho}(t)$ is the internal trip completion rate of route $\rho$ in subregion $i$; $\hat{m}_{i \to \rho(+i)}^{od,\rho}(t)$ denotes the actual transfer traffic from subregion $i$ to the next node $\rho(+i)$; and $y_{\rho(-i) \to i}^{od,\rho}(t)$ denotes the actual transfer traffic from the previous note $\rho(-i)$ to subregion $i$. If note $\rho(-i)$ is a subregion, $y_{\rho(-i) \to i}^{od,\rho}(t) = \hat{m}_{\rho(-i) \to i}^{od,\rho}(t)$; if note $\rho(-i)$ is an expressway, $y_{\rho(-i) \to i}^{od,\rho}(t)$ is equal to the flow entering subregion $i$ from the off-ramp between them, which will be presented in Section 3.2.

Specifically, Eq. (2) defines the rate of change in accumulation $n_i^{od,\rho}(t)$. In the case of (i), route $\rho$ includes only one node, and the rate is simply the exogenous demand minus the trip completion rate, which is not bounded by any capacity function. In case (ii), the current subregion $i$ is the origin and not the destination; the rate is the exogenous demand minus the transfer flow to the next node in route $\rho$. In case (iii), the current subregion $i$ is the destination and not the origin; the rate is defined as the transfer flow from the previous note minus the trip completion rate without any limitation of capacity function. In (iv), for other cases, the rate equals the transfer flow from the previous note minus the transfer flow to the next note.

If the next node $\rho(+i)$ is a subregion $j$, the external transfer flow from subregion $i$ to subregion $j \in \alpha_i$ is the minimum of the (i) trip completion rate $m_{i \to j}^{od,p}(t)$, (ii) the boundary capacity $c_{ij}$ between subregion $i$ and subregion $j$, and (iii) the fraction of receiving capacity of subregion $j$ that is assigned to $n_{i \to j}^{od,p}(t)$. The transfer flow $\hat{m}_{i \to j}^{od,p}(t)$ is estimated by the following equation:

$$\hat{m}_{i \to j}^{od,p}(t) = \min\{m_{i \to j}^{od,p}(t), c_{ij}, \frac{m_{i \to j}^{od,p}(t)}{\sum_{s_1}\sum_{s_2}\sum_{s_3 \in \alpha_j}\sum_x m_{s_3 \to j}^{s_1 s_2, x}(t) + \sum_{s_1}\sum_{s_2}\sum_{h \in \beta_j}\sum_x \delta_{E_{hj} j}^{s_1 s_2, x}(t)} \cdot c_j(n_j(t))\} \quad (3)$$

where $m_{s_3 \to j}^{s_1 s_2, x}(t)$ is the trip completion rate from the previous subregion $s_3$ to the subregion $j$ at time $t$ for vehicles with origin subregion $s_1$, destination subregion $s_2$, and route $x$. $\delta_{E_{hj} j}^{s_1 s_2, x}(t)$ is the traffic demand from off-ramp $r_{E_{hj}}$ to subregion $j$. $c_j(n_j(t))$ is the receiving capacity of subregion $j$ when the number of vehicles is $n_j(t)$, defined as:

$$c_j(n_j(t)) = c_{j,\max} \cdot (1 - \frac{n_j(t)}{n_{j,\max}}) \quad (4)$$

where $n_{j,\max}$ is the jammed accumulation of subregion $j$, and $c_{j,\max}$ is the maximum receiving capacity of subregion $j$.

If the next node $\rho(+i)$ is an expressway $E_{ij}$, the external transfer flow from subregion $i$ to expressway $E_{ij}$ is the minimum of the (i) trip completion rate $m_{i \to E_{ij}}^{od,\rho}(t)$, (ii) the capacity $C_r$ of on-ramp $r_{iE_{ij}}$, and (iii) the fraction of receiving capacity of on-ramp $r_{iE_{ij}}$ that is assigned to $n_{i \to E_{ij}}^{od,\rho}(t)$. Thus, the transfer flow $\hat{m}_{i \to E_{ij}}^{od,\rho}(t)$ is estimated as:

$$\hat{m}_{i \to E_{ij}}^{od,\rho}(t) = \min\{m_{i \to E_{ij}}^{od,\rho}(t), C_r, \frac{m_{i \to E_{ij}}^{od,\rho}(t)}{\sum_{s_1}\sum_{s_2}\sum_x m_{i \to E_{ij}}^{s_1 s_2, x}(t)} \cdot \zeta_{iE_{ij}}(t)\} \quad (5)$$

where $\zeta_{iE_{ij}}(t)$ is the receiving capacity of on-ramp $r_{iE_{ij}}$, which will be estimated through CTM theory in Section 3.2.

## 3.2 Expressway traffic model

CTM is employed to model the expressway system to predict the traffic density, outflow, and average speed on the mainline and the ramps. The cell division of the expressway is shown in Fig. 3. Based on the CTM theory, the expressway $E_{ij}$ is divided into $L_{ij}$ cells along the traveling direction, each of length $L_s$. For simplicity, each on-ramp and off-ramp is a cell of length $L_s$. When there are expressways $E_{hi}$ and $E_{ij}$ intersecting in subregion $i$, a connecting ramp $r_{E_{hi}E_{ij}}$ will be set up and any connecting ramp is assumed to be a unit cell.

The expressway is assumed to be homogeneous, and all mainline and ramp cells have known triangular fundamental diagrams, respectively. For the mainline cells, the free flow speed is $V_{fm}$; the critical density is $K_{cm}$; the capacity is $C_m$; the congested density is $K_{jm}$; and the congestion wave speed is $\omega_m$. For all of the ramp cells, the free flow speed is $V_{fr}$; the critical density is $K_{cr}$; the capacity is $C_r$; the congested density is $K_{jr}$; and the congestion wave speed is $\omega_r$. Denote $s_{E_{ij},l_{ij}}$ as a mainline cell on the expressways $E_{ij}$, $l_{ij} \in [1, L_{ij}]$.

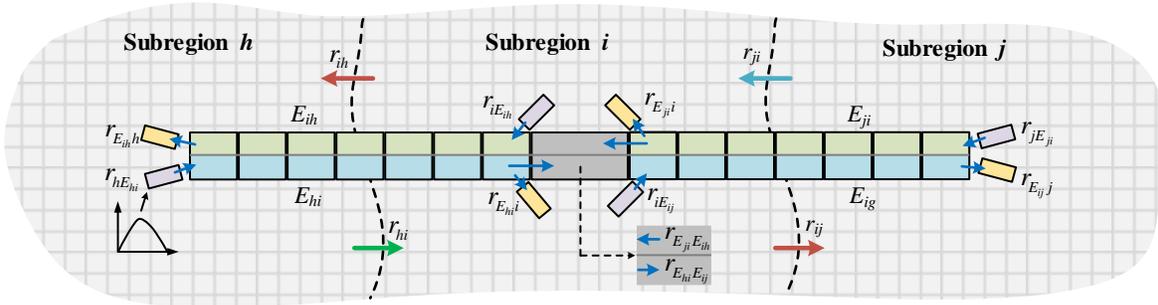

Fig. 3 Schematic diagram of expressway division in a mixed road network.

Researchers have developed multi-class CTMs to distinguish between different classes of vehicles traveling on the same road system. Depending on the modeling goals, vehicle classes can refer to different types of vehicles or driver-specific characteristics, such as driving behavior, trip purpose, etc. (Ferrara et al., 2018). This paper uses multi-class CTMs to distinguish and track traffic flows with different routes within the cells of the expressway system, aiming for more accurate traffic flow prediction. At time $t$, the traffic flow and density of the mainline cell $s_{E_{ij},l_{ij}}$ are $Q_{E_{ij},l_{ij}}(t)$ and $K_{E_{ij},l_{ij}}(t)$, respectively. Let $K^{od,\rho}_{E_{ij},l_{ij} \to p(+E_{ij})}(t)$ and $Q^{od,\rho}_{E_{ij},l_{ij} \to p(+E_{ij})}(t)$ denote the density and flow rate of vehicles in cell $s_{E_{ij},l_{ij}}$ at time $t$ with origin subregion $o$, destination subregion $d$, and next node $\rho(+E_{ij})$. Then, there is $K_{E_{ij},l_{ij}}(t) = \sum_o \sum_d \sum_\rho K^{od,\rho}_{E_{ij},l_{ij} \to p(+E_{ij})}(t)$. The flow and density of the on-ramp cell $r_{iE_{ij}}$ are denoted by $Q_{iE_{ij}}(t)$ and $K_{iE_{ij}}(t)$, respectively; the flow and density of the off-ramp $r_{E_{ij}j}$ are denoted by $Q_{E_{ij}j}(t)$ and $K_{E_{ij}j}(t)$, respectively; the flow and density of connecting ramp $r_{E_{hi}E_{ij}}$ are denoted by $Q_{E_{hi}E_{ij}}(t)$ and $K_{E_{hi}E_{ij}}(t)$, respectively. Similarly, let $K^{od,\rho}_{iE_{ij}}(t)$, $K^{od,\rho}_{E_{ij}j}(t)$ and $K^{od,\rho}_{E_{hi}E_{ij}}(t)$ denote the traffic densities within cells $r_{iE_{ij}}$, $r_{E_{ij}j}$ and $r_{E_{hi}E_{ij}}$ at time $t$ with origin subregion $o$, destination subregion $d$, and route $\rho$.

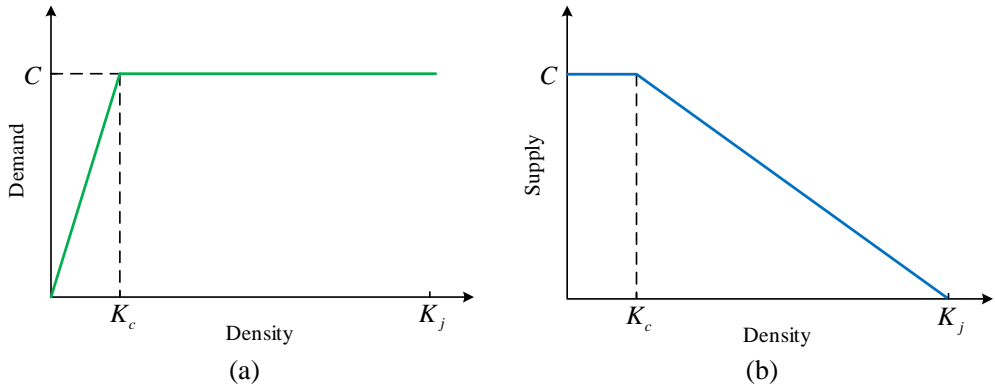

(a)     (b)

Fig. 4 (a) Demand function and (b) supply function of cells.

The theoretical flow rate of a cell is the minimum of (i) the demand of the cell and (ii) the supply capacity of its downstream cell. The traffic demand within a cell is the product of the free-flow speed and its density, and Fig. 4(a) and (b) illustrate the demand function and supply function for a cell under the definition of CTM. Since the expressway system consists of mainlines, on-ramps, off-ramps, and connecting ramps, the density dynamics of these cells are established separately in the order of entering the expressway to exiting the expressway, introduced as follows.

(a) **On-ramps.** Vehicles leaving the subregion for the expressway are the first to enter the on-ramps. If an expressway $E_{ij}$ exists, the traffic demand for traffic density $K_{iE_{ij}}^{od,\rho}(t)$ within the on-ramp $r_{iE_{ij}}$ is:

$$\delta_{iE_{ij}}^{od,\rho}(t) = \min\{V_{fr} \cdot K_{iE_{ij}}^{od,\rho}(t), C_r\} \tag{6}$$

The supply capacity of the on-ramp $r_{iE_{ij}}$ is:

$$\zeta_{iE_{ij}}(t) = \min\{\omega_r \cdot (K_{jr} - K_{iE_{ij}}(t)), C_r\} \tag{7}$$

Therefore, the traffic flow merging from the on-ramp $r_{iE_{ij}}$ into the mainline is:

$$Q_{iE_{ij}}^{od,\rho}(t) = \min\{\delta_{iE_{ij}}^{od,\rho}(k), C_r, \frac{\delta_{iE_{ij}}^{od,\rho}(t)}{\sum_{s_1}\sum_{s_2}\sum_{x}\delta_{iE_{ij}}^{s_1s_2,x}(t) + \sum_{s_1}\sum_{s_2}\sum_{x}\sum_{h\in\beta_i, i\neq j}\delta_{E_{hi}E_{ij}}^{s_1s_2,x}(t)} \cdot \zeta_{E_{ij},1}(t)\} \tag{8}$$

where $\delta_{E_{hi}E_{ij}}^{s_1s_2,x}(t)$ is the traffic flow in the connecting ramp $r_{E_{hi}E_{ij}}$ with the origin subregion $o$, destination subregion $d$, and route $x$ when there is an expressway $E_{hi}$ and expressway $E_{ij}$; $\zeta_{E_{ij},1}(t)$ is the supply capacity of the first mainline cell of expressway $E_{ij}$. The actual transfer flow of traffic demand $\delta_{E_{hi}E_{ij}}^{s_1s_2,x}(t)$ subject to constraints is denoted as $Q_{E_{hi}E_{ij}}^{s_1s_2,x}(t)$, which is calculated regarding Eq. (6). The expressway receives upstream traffic demand at the first cell of the mainline. The demand comes from both the on-ramp and the connected expressways; hence, the denominator in Eq. (8) is the sum of both.

The dynamics of a cell are due to the difference between its upstream inflow and outflow. According to the principle of accumulation conservation, the dynamics of the traffic density on on-ramp $r_{iE_{ij}}$ are:

$$\frac{dK_{iE_{ij}}^{od,\rho}(t)}{d(t)} = \hat{m}_{i\to E_{ij}}^{od,\rho}(t) - Q_{iE_{ij}}^{od,\rho}(t) \tag{9}$$

(b) **Mainline.** Vehicles in the mainline of the expressway transfer in the travel direction to the downstream cells. The transmission flow in the mainline cell $s_{E_{ij},l_{ij}}$ at time $t$ with origin subregion $o$, destination subregion $d$, and route $\rho$ is:

$$Q_{E_{ij},l_{ij}\to\rho(+E_{ij})}^{od,\rho}(t) = \min\{\delta_{E_{ij},l_{ij}\to\rho(+E_{ij})}^{od,\rho}(t), C_m, \frac{\delta_{E_{ij},l_{ij}\to\rho(+E_{ij})}^{od,\rho}(t)}{\sum_{s_1}\sum_{s_2}\sum_{x}\delta_{E_{ij},l_{ij}\to x(+E_{ij})}^{s_1s_2,x}(t)} \cdot \zeta_{E_{ij},l_{ij}+1}(t)\} \tag{10}$$

where $\delta_{E_{ij},l_{ij}\to\rho(+E_{ij})}^{od,\rho}(t)$ is the traffic demand of vehicles in cell $s_{E_{ij},l_{ij}}$ at time $t$ with origin subregion $o$, destination subregion $d$, and next node $\rho(+E_{ij})$.

The expressway mainline cell $r_{E_{ij},l_{ij}}$ updates the traffic density based on its location and the type of nodes before and after it:

$$\frac{dK^{od,p}_{E_{ij},l_{ij}\to p(+E_{ij})}(t)}{d(t)} = \begin{cases} Q^{od,p}_{iE_{ij}}(t) - Q^{od,p}_{E_{ij},l_{ij}\to p(+E_{ij})}(t) & \text{if } l_{ij}=1 \ \& \ p(-E_{ij}) = i \\ Q^{od,p}_{E_{hi}E_{ij}}(t) - Q^{od,p}_{E_{ij},l_{ij}\to p(+E_{ij})}(t) & \text{if } l_{ij}=1 \ \& \ p(-E_{ij}) = E_{hi} \\ Q^{od,p}_{E_{ij},l_{ij}-1\to j}(t) - f^{od,p}_{E_{ij}\to j}(t) & \text{if } l_{ij}=L_{ij} \ \& \ p(+E_{ij}) = j \\ Q^{od,p}_{E_{ij},l_{ij}-1\to E_{jg}}(t) - f^{od,p}_{E_{ij}\to E_{jg}}(t) & \text{if } l_{ij}=L_{ij} \ \& \ p(+E_{ij}) = E_{jg}, g \in \beta_j \\ Q^{od,p}_{E_{ij},l_{ij}-1\to p(+E_{ij})}(t) - Q^{od,p}_{E_{ij},l_{ij}\to p(+E_{ij})}(t) & \text{else} \end{cases} \quad (11)$$

where $f^{od,p}_{E_{ij}\to j}(t)$ and $f^{od,p}_{E_{ij}\to E_{jg}}(t)$ (if expressway $E_{jg}$ exists) are the transfer flow from the last cell of expressway $E_{ij}$ to the off-ramp $r_{E_{ij}j}$ and connecting ramp $r_{E_{ij}E_{jg}}$, respectively.

(c) **Off-ramps.** When vehicles travel on the expressway's mainline to reach its end, they access the connecting subregion via the off-ramp. The transfer flow $f^{od,\rho}_{E_{hi}\to i}(t)$ from expressway $E_{hi}$ to the off-ramp $r_{E_{hi}i}$ (then entering the subregion $i$) is obtained by Eq. (12):

$$f^{od,\rho}_{E_{hi}\to i}(t) = \min\{\delta^{od,\rho}_{E_{hi},L_{hi}\to i}(t), C_r, \frac{\delta^{od,\rho}_{E_{hi},L_{hi}\to i}(t)}{\sum_{s_1}\sum_{s_2}\sum_x \delta^{s_1s_2,x}_{E_{hi},L_{hi}\to i}(k)} \cdot \zeta_{E_{hi}i}(t)\} \quad (12)$$

where $\delta^{od,\rho}_{E_{hi},L_{hi}\to i}(t)$ is the traffic demand of vehicles from mainline cell $s_{E_{hi},L_{hi}}$ to off-ramp $r_{E_{hi}i}$ at time $t$ with origin subregion $o$, destination subregion $d$, and route $\rho$; $\zeta_{E_{hi}i}(t)$ is the supply capacity of the off-ramp $r_{E_{hi}i}$. Eq. (12) shows that the traffic demand from the expressway into the subregion comes from the end cell.

For vehicles exiting the off-ramp, the next node is the connected subregion, and the traffic flow from the off-ramp $r_{E_{hi}i}$ into subregion $i$ is calculated as:

$$Q^{od,\rho}_{E_{hi}i}(t) = \min\{\delta^{od,\rho}_{E_{hi}i}(t), C_r, \frac{\delta^{od,\rho}_{E_{hi}i}(t)}{\sum_{s_1}\sum_{s_2}\sum_{s_3}\sum_x m^{s_1s_2,x}_{s_3\to i}(t) + \sum_{s_1}\sum_{s_2}\sum_{g\in\beta_i}\sum_x \delta^{s_1s_2,x}_{E_{gi}i}(t)} \cdot c_i(n_i(t))\} \quad (13)$$

Therefore, the traffic density dynamic of off-ramp $r_{E_{hi}i}$ is:

$$\frac{dK^{od,\rho}_{E_{hi}i}(t)}{d(t)} = f^{od,\rho}_{E_{hi}\to i}(t) - Q^{od,\rho}_{E_{hi}i}(t) \quad (14)$$

(d) **Connecting ramps.** Furthermore, if connecting expressways are available for vehicles on the expressway, they can transfer to the connecting expressways via connecting ramps. The transfer flow $f^{od,\rho}_{E_{hi}\to E_{ij}}(t)$ from expressway $E_{hi}$ to the connecting ramp $r_{E_{hi}E_{ij}}$ (and then entering the expressway $E_{ij}$) is obtained through Eq. (15):

$$f^{od,\rho}_{E_{hi}\to E_{ij}}(t) = \min\{\delta^{od,\rho}_{E_{hi},L_{hi}\to E_{ij}}(t), C_r, \frac{\delta^{od,\rho}_{E_{hi},L_{hi}\to E_{ij}}(t)}{\sum_{s_1}\sum_{s_2}\sum_x \delta^{s_1s_2,x}_{E_{hi},L_{hi}\to E_{ij}}(k)} \cdot \zeta_{E_{hi}E_{ij}}(t)\} \quad (15)$$

where $\delta^{od,\rho}_{E_{hi},L_{hi}\to E_{ij}}(t)$ is the traffic demand of vehicles from cell $s_{E_{hi},L_{hi}}$ to connecting ramp $r_{E_{hi}E_{ij}}$ at time $t$ with origin subregion $o$, destination subregion $d$, and route $\rho$; $\zeta_{E_{hi}E_{ij}}(t)$ is the supply capacity of the connecting ramp $r_{E_{hi}E_{ij}}$. The traffic demand for connecting ramp $r_{E_{hi}E_{ij}}$ also comes from the last cell of the upstream expressway $E_{hi}$.

Finally, the traffic density dynamic of the connecting ramp $r_{E_{hi}E_{ij}}$ is:

$$\frac{dK_{E_{hi}E_{ij}}^{od,\rho}(t)}{d(t)} = f_{E_{hi} \to E_{ij}}^{od,\rho}(t) - Q_{E_{hi}E_{ij}}^{od,\rho}(t) \qquad (16)$$

where $Q_{E_{hi}E_{ij}}^{od,\rho}(t)$ is the traffic flow of vehicles from connecting ramp $r_{E_{hi}E_{ij}}$ to expressway $E_{ij}$ at time $t$ with origin subregion $o$, destination subregion $d$, and route $\rho$.

### 3.3. Route choice model for mixed road network

This section establishes a route choice model based on the mixed road network traffic model. Let $W$ denote the set of all subregion OD pairs, $\Theta_w$ denote the set of all routes between any OD pair $w \in W$. Any route in $\Theta_w$ can be denoted as $\rho(w) = \{\rho_1, \rho_2, ..., \rho_{k_\rho}\}$. Since each route is a sequence of nodes from the origin subregion to the destination subregion, the total travel time for a route is the sum of travel times at each node. Modeling with MFD (or CTM) assumes that all vehicles traveling simultaneously in a certain subregion (or a cell) have the same speed. Therefore, the travel time within a subregion is calculated by the average trip length and speed. The travel time within a cell can be extrapolated based on the density of the cell using a triangular fundamental diagram (Li et al., 2016).

The travel time at each node is calculated based on the node's location and the type of its preceding node. This includes the time spent within the node itself and the time required to traverse the ramp leading to it. However, there is no additional ramp travel time if a vehicle moves from one subregion to an adjacent subregion. Therefore, if a node in route $\rho(w)$ is subregion $i$, then the travel time of subregion $i$ is:

$$\tau_i(t) = \begin{cases} \dfrac{\bar{L}_i}{v_i(n_i(t))} & \text{if } \rho(-i) \in \alpha_i \\ \dfrac{L_s}{V_{E_{hi}i}(t)} + \dfrac{\bar{L}_i}{v_i(n_i(t))} & \text{if } \rho(-i) = E_{hi} \end{cases} \qquad (17)$$

where $V_{E_{hi}i}(t)$ is the average speed of off-ramp $r_{E_{hi}i}$.

If a node in route $\rho$ is an expressway $E_{ij}$, the travel time in $E_{ij}$ is:

$$\tau_{E_{ij}}(t) = \begin{cases} \dfrac{L_s}{V_{iE_{ij}}(t)} + \sum_{l_{ij}=1}^{L_{ij}} \dfrac{L_s}{V_{E_{ij},l_{ij}}(t)} & \text{if } \rho(-E_{ij}) = i \\ \dfrac{L_s}{V_{E_{hi}E_{ij}}(t)} + \sum_{l_{ij}=1}^{L_{ij}} \dfrac{L_s}{V_{E_{ij},l_{ij}}(t)} & \text{if } \rho(-E_{ij}) = E_{hi} \end{cases} \qquad (18)$$

where $V_{iE_{ij}}(t)$ and $V_{E_{hi}E_{ij}}(t)$ are the average speed of on-ramp $r_{iE_{ij}}$ and the connecting ramp $r_{E_{hi}E_{ij}}$, respectively; $V_{E_{ij},l_{ij}}(t)$ is the average speed of the mainline cell $s_{E_{ij},l_{ij}}$.

When travelers have more than one alternative route, they generally try to choose the route with the shortest travel time. To relax the assumption that travelers are fully informed about traffic, this paper employs a logit model to describe the driver route choice decision based on the more realistic concept of SUE (Hosseinzadeh et al., 2023). The approach involves identifying alternative routes based on origin-destination pairs, estimating the travel impedance for each route on the arterial network or expressway, and determining the choice probability for each route using a logit model (Prashker and Bekhor, 2004). The probability that the newly generated traffic demand on OD $w$ chooses route $\rho(w)$ is:

$$\theta_{\rho(w)}(t) = \frac{\exp(-\mu\tau_{\rho(w)}(t))}{\sum_{x \in \Phi_w} \exp(-\mu\tau_{x(w)}(t))} \qquad (19)$$

where $\tau_{\rho(w)}(t)$ and $\tau_{x(w)}(t)$ denote the time cost spent on routes $\rho(w)$ and $x(w)$, respectively, for the newly generated traffic flow on OD $w$; $\theta_{\rho(w)}(t)$ is the drivers' stochastic routing probability based on current network states; and $\mu$ is the logit model parameter that indicates the drivers' knowledge of the mixed road network travel time. Thus, a higher $\mu$ corresponds to a higher knowledge of the network travel time.

## 4 Expressway network design problem

By constructing the mixed road network traffic model and the route choice model, the lower-layer traffic assignment problem, as shown in Eqs. (1)-(19), is addressed. Building upon this foundation, this section tackles the upper-layer network design problem, which involves planning objectives, constraints, and decision-making methods. It operates under the assumption that the decision-maker can anticipate travelers' route choices, and they design the expressway network accordingly.

The optimization objective of the ENDP is to determine new expressways between which subregions to minimize network users' travel time while adhering to budget constraints. Minimizing the total time spent (TTS) of vehicles in the mixed road network during the simulation period is selected as the objective function, described as:

$$\min T_s \sum_{t=0}^{T-1} \left( \sum_{i \in \mathbb{R}} n_i(t) + \sum_{i \in \mathbb{R}} \sum_{j \in \mathbb{R}} \varepsilon_{ij} L_s \left[ \sum_{l_{ij}=1}^{L_{ij}} K_{E_{ij}, l_{ij}}(t) + K_{iE_{ij}}(t) + K_{E_{ij}j}(t) + \sum_{h \in \mathbb{R}} \varepsilon_{hi} \cdot K_{E_{hi}E_{ij}}(t) \right] \right) \quad (20)$$

$$\text{s.t. } \varepsilon_{ij} \in \{0,1\} \quad (21)$$

$$\sum_{i \in \mathbb{R}} \sum_{j \in \mathbb{R}} \varepsilon_{ij} \cdot \lambda_{ij} \leq \Lambda \quad (22)$$

$$\varepsilon_{ij} = \varepsilon_{ji}, \lambda_{ij} = \lambda_{ji} \quad (23)$$

$$0 \leq n_i(t) \leq n_{i,\max} \quad (24)$$

where $T_s$ is the length of time step, $T$ is the number of simulation time steps, and $\varepsilon_{ij}$ is a 0/1 decision variable. When $\varepsilon_{ij}$ is equal to 1, it indicates that there exists an expressway $E_{ij}$ and also the associated on-ramp $r_{iE_{ij}}$ and off-ramp $r_{E_{ij}j}$. The expressway is all bidirectional, hence $\varepsilon_{ij} = \varepsilon_{ji}$. $\lambda_{ij}$ denotes the construction cost of the expressway $E_{ij}$. Assuming the construction costs of both upstream and downstream ramps have been considered, the construction costs for both directions of an expressway are equivalent. $\Lambda$ denotes the total construction budget, which should be accounted for as a cost constraint in the expressway network design.

It should be noted that this study assumes the lengths, lane widths, and number of lanes of the candidate expressways are known. The ENDP only considers whether a new expressway should be constructed, making it a discrete network design problem. While the branch-and-bound method and mathematical programming can offer optimal solutions, they are inefficient for large networks due to extensive computational requirements (Farahani et al., 2013). Conversely, metaheuristic methods like simulated annealing, genetic algorithms, etc., are not reliant on specific mathematical properties and can find near-global optimum solutions more swiftly (Meng and Yang, 2002; Mathew and Sharma, 2009). These methods excel in tackling complex nonconvex problems with numerous decision variables. Therefore, the particle swarm algorithm from the suite of metaheuristic algorithms has been selected for problem-solving.

## 5 Case Studies

Based on the methodology introduced above, this section explores the expressway design problem for five subregions using the scenario depicted in Fig. 1(b) of Section 2.1 as a case study. Each subregion is assigned a distinct and well-defined MFD function. We use the MFD function developed by Geroliminis and Daganzo (2008) from the data of Yokohama as the base function and adjust it into five functions by varying the average trip length. Table 2 presents the MFD functions for each subregion. The maximum vehicle capacity per subregion is set at 10,000 vehicles. There are 8 candidate expressways in Fig. 2, and the mainline length of each candidate expressway is shown in Table 3. In applying CTM to the expressway road segments, the length of each cell is $L_s = 500$ m. The mainline of the candidate expressway is assumed to have a free-flow speed of 80 km/h and a capacity of 6000 veh/h, while all ramps have a free-flow speed of 40 km/h with a capacity of 3000 veh/h. Assuming a construction cost of \$5 million per kilometer for both directions of the expressway, the construction cost for each candidate expressway can be determined.

Table 2. MFD function and average trip length of each subregion.

| Subregion | Production $P(n)$ (veh·m/s) | Average trip length (m) |
|---|---|---|
| 1 | $1.4877e^{-7}n^3 - 2.9815e^{-3}n^2 + 15n$ | 4800 |
| 2 | | 5200 |

|   |   |
|---|---|
| 3 | 4700 |
| 4 | 5500 |
| 5 | 4500 |

Table 3. The mainline length of each candidate expressway.

| Expressway | $E_{12}/E_{21}$ | $E_{23}/E_{32}$ | $E_{34}/E_{43}$ | $E_{14}/E_{41}$ | $E_{15}/E_{51}$ | $E_{25}/E_{52}$ | $E_{35}/E_{53}$ | $E_{45}/E_{54}$ |
|---|---|---|---|---|---|---|---|---|
| Length (m) | 6500 | 6500 | 6500 | 6000 | 5500 | 6000 | 5500 | 6000 |

### 5.1 Investment budget factors

To investigate the ENDP under various funding scenarios, we consider six total budget levels: $0M, $50M, $100M, $150M, $200M, and $250M, respectively. Each budget level corresponds to an optimal expressway design solution. In solving the optimization objective of the ENDP, the simulation duration for the experiment is set to 1 hour. Table 4 presents the OD demands between subregions in the first 30 minutes, while the traffic demands between subregions in the subsequent 30 minutes are all zero to simulate congestion formation and dissipation processes. Figs. 5(a)-(f) illustrate the decision schemes corresponding to the six investment budgets, labeled as schemes 1-6, respectively. In these figures, 'x' denotes that the variable $\varepsilon_{ij}$ is not within the decision scope of the expressway design in this study; '0' indicates that the decision scheme suggests not setting up the expressway; and '1' indicates that the decision scheme recommends establishing the expressway.

Fig. 5(a) illustrates the absence of any expressway construction. In Fig. 5(b), expressway $E_{34}(E_{43})$ takes priority for construction with a $50M funding input. As depicted in Fig. 5(c), expressways $E_{12}(E_{21})$, $E_{23}(E_{32})$, and $E_{34}(E_{43})$ are prioritized for construction with a $100M budget. Moving to Fig. 5(d), expressways $E_{34}(E_{43})$, $E_{15}(E_{51})$, $E_{25}(E_{52})$, $E_{35}(E_{53})$, and $E_{45}(E_{54})$ are prioritized for construction when the budget reaches $150M. Fig. 5(e) indicates that, with a $200M budget, constructing expressway $E_{12}(E_{21})$ becomes a priority based on the decision scheme established under the $150M budget. Lastly, Fig. 5(f) suggests that with an investment exceeding $250M, constructing expressways between all subregions may be considered to alleviate overall travel time.

Table 4 Traffic demand between subregion ODs (veh/h).

|   | 1 | 2 | 3 | 4 | 5 |
|---|---|---|---|---|---|
| 1 | 3400 | 3200 | 3000 | 3200 | 3400 |
| 2 | 3200 | 3400 | 3200 | 3000 | 3000 |
| 3 | 3000 | 3200 | 3400 | 3200 | 3400 |
| 4 | 3200 | 3000 | 3200 | 3400 | 3000 |
| 5 | 3400 | 3000 | 3400 | 3000 | 3400 |

| $\varepsilon_{ij}$ | 1 | 2 | 3 | 4 | 5 |
|---|---|---|---|---|---|
| 1 | x | 0 | x | 0 | 0 |
| 2 | 0 | x | 0 | x | 0 |
| 3 | x | 0 | x | 0 | 0 |
| 4 | 0 | x | 0 | x | 0 |
| 5 | 0 | 0 | 0 | 0 | x |

(a) Scheme 1: Λ = $0M

| $\varepsilon_{ij}$ | 1 | 2 | 3 | 4 | 5 |
|---|---|---|---|---|---|
| 1 | x | 0 | x | 0 | 0 |
| 2 | 0 | x | 0 | x | 0 |
| 3 | x | 0 | x | 1 | 0 |
| 4 | 0 | x | 1 | x | 0 |
| 5 | 0 | 0 | 0 | 0 | x |

(b) Scheme 2: Λ = $50M

| $\varepsilon_{ij}$ | 1 | 2 | 3 | 4 | 5 |
|---|---|---|---|---|---|
| 1 | x | 1 | x | 0 | 0 |
| 2 | 1 | x | 1 | x | 0 |
| 3 | x | 1 | x | 1 | 0 |
| 4 | 0 | x | 1 | x | 0 |
| 5 | 0 | 0 | 0 | 0 | x |

(c) Scheme 3: Λ = $100M

| $\mathcal{E}_{ij}$ | 1 | 2 | 3 | 4 | 5 |
|---|---|---|---|---|---|
| 1 | x | 0 | x | 0 | 1 |
| 2 | 0 | x | 0 | x | 1 |
| 3 | x | 0 | x | 1 | 1 |
| 4 | 0 | x | 1 | x | 1 |
| 5 | 1 | 1 | 1 | 1 | x |

(d) Scheme 4: Λ = $150M

| $\mathcal{E}_{ij}$ | 1 | 2 | 3 | 4 | 5 |
|---|---|---|---|---|---|
| 1 | x | 1 | x | 0 | 1 |
| 2 | 1 | x | 0 | x | 1 |
| 3 | x | 0 | x | 1 | 1 |
| 4 | 0 | x | 1 | x | 1 |
| 5 | 1 | 1 | 1 | 1 | x |

(e) Scheme 5: Λ = $200M

| $\mathcal{E}_{ij}$ | 1 | 2 | 3 | 4 | 5 |
|---|---|---|---|---|---|
| 1 | x | 1 | x | 1 | 1 |
| 2 | 1 | x | 1 | x | 1 |
| 3 | x | 1 | x | 1 | 1 |
| 4 | 1 | x | 1 | x | 1 |
| 5 | 1 | 1 | 1 | 1 | x |

(f) Scheme 6: Λ = $250M

Fig. 5 Decision schemes with budgets of (a) $0M, (b) $50M, (c) $100M, (d) $150M, (e) $200M, and (f) $250M.

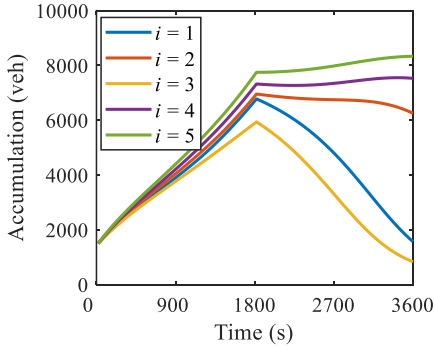
(a) Scheme 1

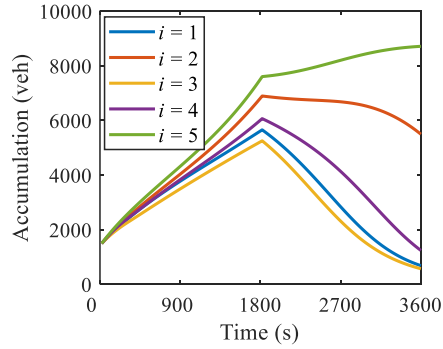
(b) Scheme 2

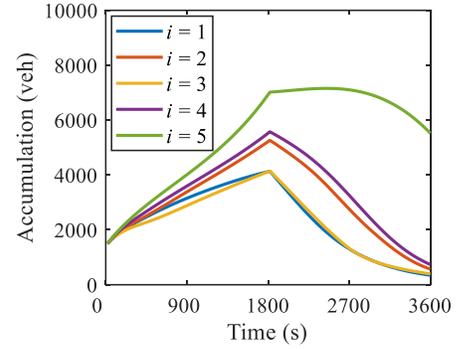
(c) Scheme 3

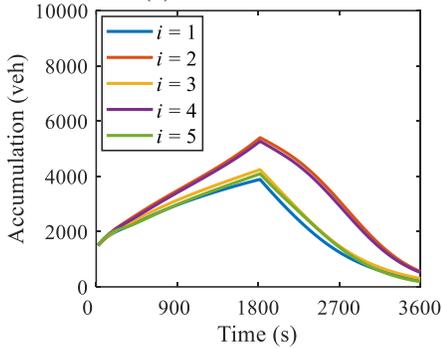
(d) Scheme 4

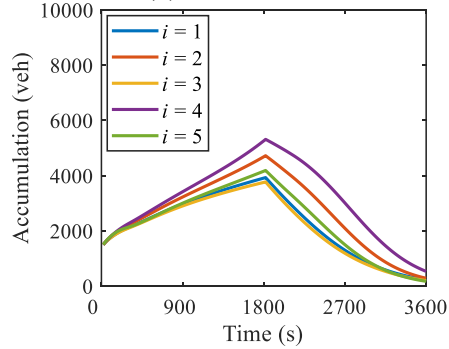
(e) Scheme 5

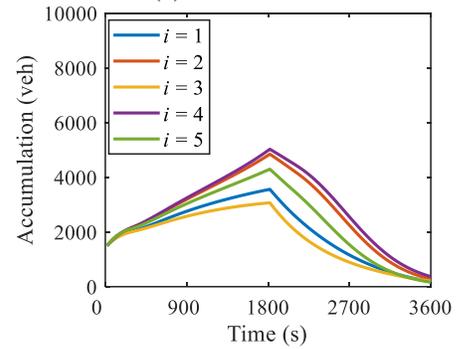
(f) Scheme 6

Fig. 6 Number of vehicles in subregions under different decision schemes.

Fig. 6 shows the trend of vehicles in every subregion for the six decision schemes in Fig. 5. In Scheme 1, where no expressways are designed, the average number of vehicles for all subregions during the simulation period is 25,299 veh. Scheme 2 designs expressways between Subregion 3 and Subregion 4, leading to a notable decrease in vehicle count in both subregions, resulting in an average count of 21,822 vehicles across all subregions, marking a 13.74% decrease from Scheme 1. Scheme 3 further designs expressways between Subregion 1 and Subregion 2, and Subregion 2 and Subregion 3, so the number of vehicles in Subregion 2 has decreased significantly. The average number of vehicles for all subregions under Scheme 3 is 16,605 veh, a decrease of 23.91% compared to Scheme 2. Subregion 5 remains relatively high in vehicle count as no expressway links it with other subregions in Schemes 1-3. From scheme 4 onwards, it is prioritized to build expressways between subregion 5 and other subregions. The average number of vehicles in all of the subregions under scheme 4 is 13,352 veh, a decrease of 19.59% compared to scheme 3. With the increase in budget, more expressways could be constructed. The average number of vehicles for all subregions under schemes 5 and 6 are 12,620 veh and 11,949 veh, respectively, which are not much different from the results of scheme 4.

The results depicted in Fig. 6 demonstrate that constructing expressways between subregions alleviates congestion within these areas. However, it is important to note that some vehicles are diverted to the expressways. Fig. 7 illustrates the trend of vehicle counts in (a) all subregions, (b) expressways, and (c) the entire mixed road network across different decision schemes. The findings indicate that the average number of vehicles within the subregions gradually decreases as the scale of expressways between subregions increases. Furthermore, introducing new expressways attracts more vehicles from subregions to the expressway network, leading to a progressive increase in vehicle count within the expressway system.

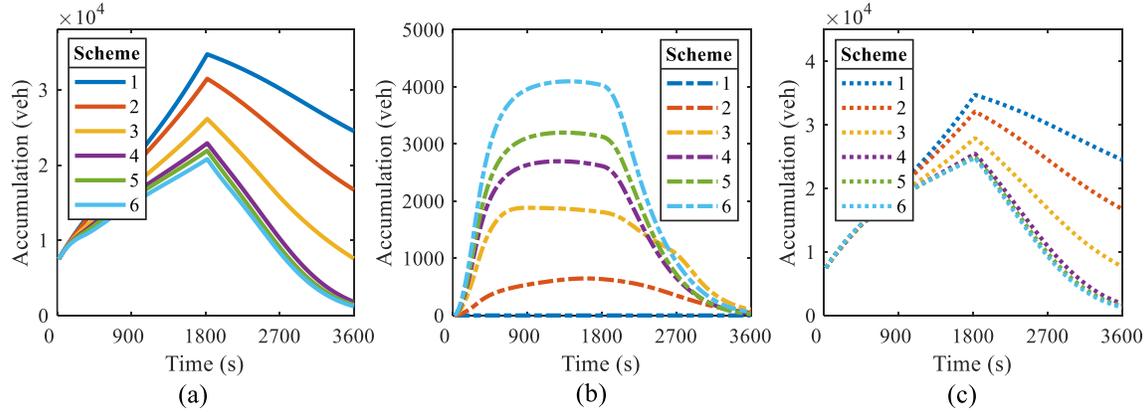

Fig. 7 The number of vehicles in (a) all subregions (b) expressways, and (c) the entire mixed road network under different decision schemes.

The average number of vehicles, average trip completion rate, TTS, and construction costs under different construction schemes are shown in Table 5. The magnitude of change in metrics for each scheme compared to the previous scheme was also presented. The results show that a reasonable increase in expressways can improve traffic efficiency. However, it is worth noting that further increasing the number of expressways will not significantly improve traffic conditions when the traffic demand has not fully reached a certain level. Specifically, when the construction budget is increased to $50M, the average number of vehicles is reduced by 12.14% compared to the situation without the expressway; when the construction budget is raised to $100M, the average number of vehicles is further reduced by 19.66% compared to the scheme with a budget of $50M. However, when the budget is raised to $200M, the average number of vehicles is reduced by only 3.08% compared to the scheme with a budget of $150M. The results indicate that when the traffic demand is insufficient, continuing to build more expressways has a diminishing marginal effect on improving traffic efficiency.

Table 5 The comparison of evaluation metrics and costs under different decision schemes.

| Scheme | Average accumulations (veh) | Average travel completion flow (veh/h) | TTT (veh·h) | Construction cost (dollar) |
|---|---|---|---|---|
| 1 | 25299 (-) | 23883 (-) | 11245 (-) | 0 |
| 2 | 22227 (↓12.14%) | 28755 (↑20.40%) | 10470 (↓6.89%) | 32.5M |
| 3 | 17858 (↓19.66%) | 34122 (↑18.66%) | 9221 (↓11.93%) | 97.5M |
| 4 | 14919 (↓16.46%) | 39876 (↑16.86%) | 8452 (↓8.34%) | 147.5M |
| 5 | 14460 (↓3.08%) | 40263 (↑0.97%) | 8322 (↓1.54%) | 180M |
| 6 | 14294 (↓1.15%) | 40784 (↑1.29%) | 8293 (↓0.35%) | 242.5M |

### 5.2 Traffic demand factors

When planning expressways, it is crucial to closely align the design scheme with actual traffic demands. Under the same construction budget, varying traffic demands between subregions may yield different expressway design outcomes. Similarly, the budget is divided into six levels of $0M, $50M, $100M, $150M, $200M and $250M, and the traffic demand in Table 4 is replaced by the traffic demand in Table 6 for decision making, and the results are shown in Fig. 8. Notably, the new decision results are the same as the previous decision results (shown in Fig. 5) for the $0M, $50M and $250M budget scenarios. However, there are significant differences in the optimal planning schemes for budgets of $100M, $150M and $200M. Therefore, the planning and design of expressways should be flexibly adjusted for different traffic demand distributions, ensuring efficient resource utilization and maximizing the efficacy of the road network.

Table 6 Traffic demand between subregions ODs (veh/h).

|   | 1 | 2 | 3 | 4 | 5 |
|---|---|---|---|---|---|
| 1 | 3400 | 2600 | 2200 | 2600 | 3000 |
| 2 | 2600 | 3400 | 2600 | 3000 | 3600 |
| 3 | 2200 | 2600 | 3400 | 2600 | 3000 |

| | 4 | 2600 | 3000 | 2600 | 3400 | 3600 |
| | 5 | 3000 | 3600 | 3000 | 3600 | 3400 |

| $\mathcal{E}_{ij}$ | 1 | 2 | 3 | 4 | 5 |
|---|---|---|---|---|---|
| 1 | x | 0 | x | 0 | 0 |
| 2 | 0 | x | 0 | x | 0 |
| 3 | x | 0 | x | 0 | 0 |
| 4 | 0 | x | 0 | x | 0 |
| 5 | 0 | 0 | 0 | 0 | x |

(a) $\Lambda = \$0M$

| $\mathcal{E}_{ij}$ | 1 | 2 | 3 | 4 | 5 |
|---|---|---|---|---|---|
| 1 | x | 0 | x | 0 | 0 |
| 2 | 0 | x | 0 | x | 0 |
| 3 | x | 0 | x | 1 | 0 |
| 4 | 0 | x | 1 | x | 0 |
| 5 | 0 | 0 | 0 | 0 | x |

(b) $\Lambda = \$50M$

| $\mathcal{E}_{ij}$ | 1 | 2 | 3 | 4 | 5 |
|---|---|---|---|---|---|
| 1 | x | 0 | x | 0 | 1 |
| 2 | 0 | x | 0 | x | 1 |
| 3 | x | 0 | x | 0 | 0 |
| 4 | 0 | x | 0 | x | 1 |
| 5 | 1 | 1 | 0 | 1 | x |

(c) $\Lambda = \$100M$

| $\mathcal{E}_{ij}$ | 1 | 2 | 3 | 4 | 5 |
|---|---|---|---|---|---|
| 1 | x | 0 | x | 1 | 1 |
| 2 | 0 | x | 0 | x | 1 |
| 3 | x | 0 | x | 0 | 1 |
| 4 | 1 | x | 0 | x | 1 |
| 5 | 1 | 1 | 1 | 1 | x |

(d) $\Lambda = \$150M$

| $\mathcal{E}_{ij}$ | 1 | 2 | 3 | 4 | 5 |
|---|---|---|---|---|---|
| 1 | x | 1 | x | 1 | 1 |
| 2 | 1 | x | 0 | x | 1 |
| 3 | x | 0 | x | 0 | 1 |
| 4 | 1 | x | 0 | x | 1 |
| 5 | 1 | 1 | 1 | 1 | x |

(e) $\Lambda = \$200M$

| $\mathcal{E}_{ij}$ | 1 | 2 | 3 | 4 | 5 |
|---|---|---|---|---|---|
| 1 | x | 1 | x | 1 | 1 |
| 2 | 1 | x | 1 | x | 1 |
| 3 | x | 1 | x | 1 | 1 |
| 4 | 1 | x | 1 | x | 1 |
| 5 | 1 | 1 | 1 | 1 | x |

(f) $\Lambda = \$250M$

Fig. 8 Decision schemes with budgets of (a) $0M, (b) $50M, (c) $100M, (d) $150M, (e) $200M, and (f) $250M under traffic demand of Table 6.

## 6. Conclusion

This paper addresses the design problem of expressway networks within large-scale urban road networks, and the main work includes three aspects. Firstly, we established a traffic flow model for the mixed road network utilizing the MFD and CTM and developed a traffic assignment model at the macroscopic level. Secondly, we proposed an expressway network design model based on the traffic flow model of the mixed road network, determining design schemes under construction budget constraints to maximize investment returns. Finally, we validated the practicality and effectiveness of the proposed design model through a case study, highlighting how investment costs and variations in traffic demand between subregions directly influence decision outcomes.

Overall, this paper has two main contributions. On the one hand, compared with the traditional approaches of RNDP that take intersections as nodes of routes, this study takes every subregion and expressway as nodes of routes in a mixed road network. This innovative perspective aids in understanding and constructing the traffic flow exchange mechanism between urban subregions and expressways, offering a more macroscopic and simplified approach for predicting travel times within the road network. On the other hand, we propose an expressway network design model with multiple subregions, which can realize optimal expressway design decisions under certain budget constraints and adapt to the differentiated traffic demands among different urban subregions.

In the case studies, we investigated how the construction budget and traffic demand affect the decision outcomes of the expressway network and arrived at the following conclusions. Firstly, as the construction budget increases, the expressway network gradually expands, alleviating overall traffic congestion in the mixed road network. Particularly in the initial stages, new expressways can significantly reduce network congestion. However, as the expressway network further expands, the impact of new expressways on traffic conditions diminishes if traffic demand does not increase. Therefore, when planning expressway construction, decision-makers should consider costs and potential benefits, adjusting planning strategies based on actual traffic demand to avoid unnecessary resource waste. Secondly, variations in traffic demand between subregions result in different construction schemes. This underscores the importance of adjusting budget allocations according to specific traffic demands to ensure effective resource utilization and maximize network efficiency.

This paper's expressway network design model applies to a multi-subregion network with a well-defined topology. For the sake of model simplicity, only one ramp is provided between each expressway and the connecting subregion, and therefore, it does not consider the case of completing a trip via the expressway within the same subregion. Even though the real network is more complex and has more ramps, this simplification still preserves the impact of the expressway on the travel mode of vehicles within the subregions. Planning for ramp distribution within the same subregion will be one of the future studies.